\documentclass[times, 10pt,twocolumn]{article}
\usepackage{latex8}
\usepackage{times}

\usepackage{algorithm}
\usepackage{algorithmic}

\usepackage{cite}      

\usepackage{graphicx}  

\usepackage{amsmath}   

\begin{document}
%
\title{Defect-Tolerant CMOL Cell Assignment via Satisfiability}
%
%

\author{William N. N. Hung\\
Synplicity Inc.\\
600 W. California Ave\\
Sunnyvale, California, USA\\
william\_hung@alumni.utexas.net\\
\and
Changjian Gao, Xiaoyu Song, and Dan Hammerstrom\\
Department of Electrical \& Computer Engineering\\
Portland State University\\
Portland, Oregon, USA\\
cgao,song,strom@ece.pdx.edu\\
}

\maketitle

\thispagestyle{empty}   

\begin{abstract}
We present a CAD framework for CMOL, a hybrid CMOS/ molecular circuit architecture.
Our framework first transforms any logically synthesized circuit
based on AND/OR/NOT gates to a NOR gate circuit, and then maps
the NOR gates to CMOL.
We encode the CMOL cell assignment problem as boolean conditions.
The boolean constraint is satisfiable if and only if there is a way
to map all the NOR gates to the CMOL cells.
We further investigate various types of static defects for the CMOL
architecture, and propose a reconfiguration technique that can
deal with these defects through our CAD framework.
This is the first automated framework for CMOL cell assignment,
and the first to model several different CMOL static defects.
Empirical results show that our approach is efficient and scalable.
\end{abstract}


%

\newtheorem{theorem}{Theorem}
\newtheorem{definition}{Definition}
\newtheorem{lemma}{Lemma}

\section{Introduction}
\label{sec:intro}

In recent years, nanoelectronics has made tremendous progress,
with advances in novel nanodevices \cite{Xiang2006-Ge/Si},
nano-circuits \cite{Bachtold2001-Logic,Friedman2005-High-speed},
nano-crossbar arrays
\cite{Chen2003-Nanoscale,Kuekes2005-crossbar,Snider2004-CMOS-like},
manufacture by nanoimprint lithography
\cite{Zankovych2001-Nanoimprint,Resnick2003-Imprint}, CMOS/nano
co-design architectures
\cite{Likharev2005-CMOL:,DeHon2003-Stochastic,Ziegler2003-CMOS/nano},
and applications
\cite{Strukov2005-Prospects,Strukov2005-FPGA:,Turel2005-Architectures}.
Although a two-terminal nanowire crossbar array does not have the
functionality of FET-based circuits, it has the potential for
incredible density and low fabrication costs
\cite{Likharev2005-CMOL:}.
Likharev and his colleagues \cite{Likharev2005-CMOL:} have
developed the concept of CMOL (Cmos / nanowire / MOLecular hybrid)
as a likely implementation technology for charge-based
nanoelectronics devices. Examples include memory, FPGA, and
neuromorphic CrossNets
\cite{Strukov2005-Prospects,Strukov2005-FPGA:,Turel2005-Architectures}.

In this paper, we present a framework for CMOL cell assignment.
We transform any boolean circuit based on AND/OR/NOT gates 
to a circuit of NOR gates, and then map the NOR gates to the CMOL architecture.
We formulate the CMOL cell mapping task as a set of boolean conditions,
and solve them through satisfiability.
Prior work on CMOL \cite{Likharev2005-CMOL:} was assigning cells by hand.
Our technique is the first automated CMOL cell assignment framework.
We further investigate various defect models for the CMOL technology,
and propose a reconfiguration technique that can
deal with all these defects through our cell assignment framework.
This is the first detailed study on numerous CMOL defect models.






\section{Background about CMOL}
\label{sec:background}

\begin{figure}
\centering
\includegraphics[width=2.8in]{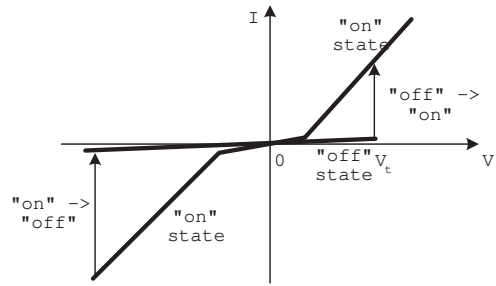}
\caption{Schematic \emph{I}-\emph{V} curve of a two-terminal
nanodevice (adapted from \cite{Likharev2005-CMOL:}).}
\label{fig:device}
\end{figure}

CMOL was originally developed by 
Likharev and his colleagues \cite{Likharev2005-CMOL:}.
The nanodevice in CMOL is a binary ``latching switch" based on
molecules with two metastable internal states. Fig.
\ref{fig:device} shows the schematic \emph{I}-\emph{V} curve of
this two-terminal nanodevice. Qualitatively, if the
drain-to-source voltage is low during programming, the nanodevice
will be in the ``off" state with a high resistance; if the applied
voltage is greater than a certian value, the nanodevice will be in
the ``on" state with a lower resistance. In the operating mode, if
the nanodevice is in the ``on" state and the applied voltage to
the drain and source is greater than the threshold voltage $V_t$,
the \emph{I}-\emph{V} curve will be like the \emph{I}-\emph{V}
curve of a finite resistor. If the applied voltage is less than
threshold voltage, then the nanodevice is virtually in the ``off"
state. However, it is not certain yet how large the on-off
resistances are and how long the nanodevice can keep its
programmed state. From our previous analysis
\cite{Gao2007-Cortical}, for an aggressive assumption of CMOL's
parameters with 6 nm nanowire pitch, the nanodevice ``on"
resistance could exhibit a higher value than that of a reasonable length
of nanowire (e.g., 6 $\mu$m). To avoid routing the critical signal
path via multiple nanodevices is one of the synthesis and routing rules 
we need to be aware of. 
In this paper, we address the CMOL cell assignment problem with
the understanding that routing between two cells through the nanowire fabric
will involve one nanodevice only (see Section~\ref{sec:routing}).

\begin{figure}
\centering
\includegraphics[width=3in]{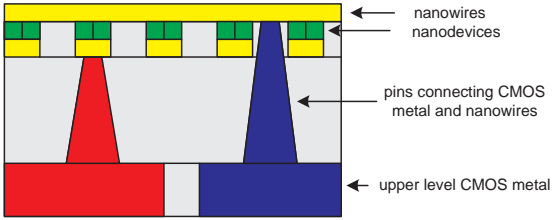}
\caption{Basic CMOL circuit. The nanowires are on top of CMOS
circuits, with pins connecting CMOS upper-level metals and
nanowires. (adapted from \cite{Strukov2005-FPGA:}).}
\label{fig:interface}
\end{figure}

Fig. \ref{fig:interface} shows the basic CMOL circuit, especially
the interface between CMOS and nanowires. The pins connect the
CMOS upper-level metals and the nanowires. The nanodevices are
sandwiched between the two levels of perpendicular nano-imprinted
nanowires. This unique structure solves the problems of addressing
much denser nanodevices with sparser CMOS components. Each
nanodevice is accessed by the two perpendicular nanowires which
connect to the nanodevice. The nanowires are in turn connected by
the pins which interface with the CMOS circuits. With $O(N)$
nanowires and pins, we could address $O(N^2)$ nanodevices.

\begin{figure}
\centering
\includegraphics[width=2in]{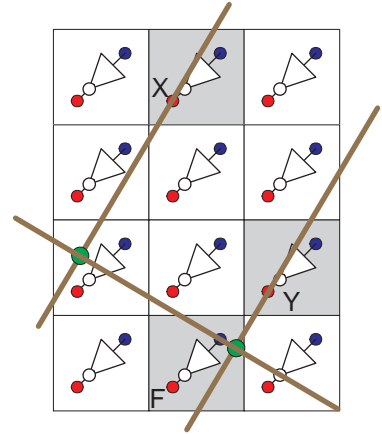}
\caption{CMOL FPGA example configuration (adapted from
\cite{Strukov2005-FPGA:}).} \label{fig:fpgalogic}
\end{figure}

Strukov and Likharev
\cite{Strukov2005-FPGA:,Strukov2006-reconfigurable} proposed the
CMOL FPGA idea to fully explore the regularity of the CMOL
architecture. Because the nanodevices are non-volatile switches,
the CMOL FPGA could program those nanodevices and route the
signals from CMOS to the nanowires and nanodevices, and back to
CMOS again. All logic functions should be done in the CMOS level.
To further explore the architectural regularity, they proposed
cell-like CMOS stucture, as shown in Fig. \ref{fig:fpgalogic}. In
each cell, there is an inverter in the CMOS level. One direction
of nanowires receive signals from the outputs of the inverters in
the CMOS level. Those nanowires are OR'ed together with another
direction of nanowire according to the nanodevice configurations
in the nanowire level. The OR'ed signal goes to the inverter's
input, which is on the CMOS level. This OR-NOT logic is the
fundamental logic of CMOL FPGA. Any combinational logics should be
expressed in the OR-NOT (or NOR) way. For example, in Fig.
\ref{fig:fpgalogic}, $X, Y, F$ are three signals connected with
the three grayed cells' output pins. With the illustrated nanowire
connections (brown lines) and ``ON" nanodevices (green dots), the logic
expression is $F=\overline{X+Y}$.

Based on Strukov and Likharev's CMOL FPGA, Snider and Williams
\cite{Snider2007-Nano/CMOS} proposed field-programmable nanowire
interconnect (FPNI) with more conservative circuit parameters than
CMOL FPGA's, such as wider nanowires and wider nanowire pitches,
sparser crossbar arrays, and larger pins. Moreover, the FPNI
assumes combination of logic gates (e.g., NAND/AND), buffers,
flipflops in the CMOS cell (or hypercell), which is similar to the
concurrent CMOS FPGA architectures. Although the area consumption
of the FPNI is larger than that of the CMOL FPGA, it shows at
least one order of magnitude of reduction in area compared with
CMOS FPGA \cite{Snider2007-Nano/CMOS}. And FPNI should be much
more practical for manufacturing than CMOL FPGA as projected in 10 years
\cite{Snider2007-Nano/CMOS}.

Strukov and Likharev \cite{Strukov2005-FPGA:} presented the CMOL FPGA
and performed cell assignment task manually for simple regular-structured
boolean circuits. They also presented a reconfigurable architecture 
\cite{Strukov2006-reconfigurable} for CMOL FPGA,
that grouped CMOL cells to form lookup-tables (LUTs),
which can utilize existing (LUT-based) FPGA CAD tools..
However, that work also did not solve the CMOL cell assignment problem.
In this paper, we solve the CMOL cell assignment problem via satisfiability
and extend it as a reconfiguration tool for various CMOL defects.


\section{CAD Framework}\label{sec:framework}

The CMOL architecture presented by \cite{Strukov2005-FPGA:}
is capable of implementing a circuit of NOR gates, as explained in 
Fig.~\ref{fig:fpgalogic}. This means the logic synthesis front-end must
present a circuit in terms of NOR gates. We can then place and route these
NOR gates on the CMOL cells.
There are many logic synthesis tools that can optimize boolean circuits
based on AND/OR/NOT gates, but there are not much recent work on optimization
of NOR gate circuits. In this section, we first present a simple algorithm
to convert any circuit of AND/OR/NOT gates into a circuit of NOR gates. 
We then describe how to do routing on CMOL cells and
present our satisfiability-based cell assignment method.
We use an adder as an example to illustrate our ideas.

\subsection{NOR Gate Transformation} \label{sec:NOR}

\begin{figure}
\centering
\includegraphics[width=2.5in]{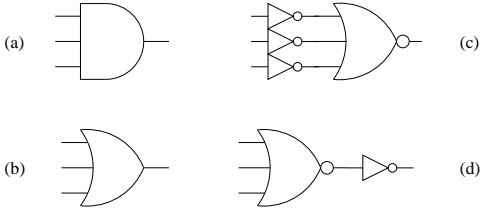}
\caption{Transform AND/OR to NOR gate}
\label{fig:demorgan}
\end{figure}

Given any boolean circuit in terms of AND/OR/NOT gates, we want to transform
the AND gates and OR gates into NOR gates 
(NOT gates are considered as single input NOR gates).
The pseudo code is shown in Algorithm~\ref{alg:NOR}.
We use De Morgan's law to convert AND/OR to NOR gate (lines 3 and 5), 
also shown in figure~\ref{fig:demorgan}.
We then remove stacked inverters in lines 8--14.
We use $fanin(u)$ to denote the set of gates that drive the input of gate $u$.
The double for loop from line 15 to line 22 will remove duplicated inverters.
Notice that the complexity of this double for loop can be reduced if we hash
all the gates based on inputs nets (i.e., performance speedup).

\begin{algorithm}
\caption{Convert AND/OR/NOT to NOR/NOT}
\label{alg:NOR}
\begin{algorithmic}[1]
\REQUIRE Input: Circuit $K$ with AND/OR/NOT gates
\ENSURE Output: Circuit $K$ with NOR/NOT gates
\FOR{each gate $g$ in $K$}
\IF{$g$ is AND gate}
\STATE convert $g$ to NOR with inverters at its inputs
\ELSIF{$g$ is OR gate}
\STATE convert $g$ to NOR with inverter at its output
\ENDIF
\ENDFOR
\FOR{each inverter $u$ in $K$}
\STATE $\{v\}$ = fanin($u$)
\IF{$v$ is inverter}
\STATE disconnect the output net $n$ of $u$
\STATE connect $n$ to the input net of $v$
\ENDIF
\ENDFOR
\FOR{each inverter $u$ in $K$}
\FOR{each inverter $v$ in $K$}
\IF{$(u \neq v) \land (fanin(u) = fanin(v))$}
\STATE disconnect the output net $n$ of $v$
\STATE connect $n$ to the input net of $u$
\ENDIF
\ENDFOR
\ENDFOR
\end{algorithmic}
\end{algorithm}

\begin{theorem}
For any boolean circuit in the Product-of-Sum (POS) format 
with at least one AND gate and all its inputs driven by OR gates,
Algorithm~\ref{alg:NOR} will result in a NOR gate circuit
with the same number of NOR gates as the number of
(AND/OR/NOT) gates in the original POS circuit.
\end{theorem}

Proof:
The AND gate will be converted into a NOR gate with inverters at its input.
Similarly, the OR gates will be converted into NOR gates 
with inverters at the output.
Since all OR gate outputs are connected to the AND gate and all AND gate
inputs are driven by OR gates, we will end up with inverters driving inverters
which can be easily eliminated. Hence the number of gates remains the same.


\subsection{Routing} \label{sec:routing}

Routing between CMOL cells is pre-determined by the nanowire fabric.
Each CMOL cell has one output nanowire and one input nanowire
which are orthogonal to each other. The input nanowires for
all CMOL cells are oriented in the same direction (parallel),
and the same property is true for all the output nanowires.
Hence, there is only one intersection between the output nanowire
of one CMOL cell and the input nanowire of another CMOL cell.
If we want to connect these two cells, we need to program
the nanodevice at the corresponding intersection to be ``ON".
For example, in figure~\ref{fig:fpgalogic}, the output nanowire
of cell $X$ and the input nanowire of cell $Y$ has a unique
intersection shown on the left side of the figure.
By turning the corresponding nanodevice ON or OFF, we can connect
or disconnect the route from $X$ to $Y$ respectively.

According to \cite{Strukov2005-FPGA:}, 
there are periodic breaks in the nanowire fabric, 
such that each input/output nanowire has a fixed length based on the period.
Hence each CMOL cell can only be connected to a limited number of neighboring
CMOL cells. The set of CMOL cells that can be connected to the input
of a particular cell $X$ is called the input {\bf connectivity domain}
of $X$. Similarly, the output connectivity domain refers to the set of
cells that can be connected to the output of $X$.

Although the input/output connections between any two CMOL cells through
the nanowire fabric is pre-determined and limited by the connectivity domain,
it is still possible for one cell to communicate with cells outside its
connectivity domain. For example, cell $A$ send its output to cell $B$
(which is within the connectivity domain of $A$), and then cell $B$
can send its output to cell $C$ (which is inside the connectivity domain of
$B$ but outside the connectivity domain of $A$). However, some physical
constraints may prevent people from using this idea for timing critical paths
\cite{Gao2007-Cortical}.
In addition, the CMOL
architecture implements the NOR logic (or NOT gate for single input case),
we may have to use two intermediate cells (i.e. two inverters) to maintain
the same logic polarity. This idea is similar to buffer insertion
in traditional ASIC CAD flow. For the rest of this paper, we assume
that these types of connecting through multiple CMOL cells is handled
by the generation of logic circuit, which can be done through logic synthesis.

\subsection{CMOL Cell Assignment} 
\label{sec:satis}



We are given a collection $C$ of CMOL cells, 
the number of cells in this collection is $||C||$. 
Each CMOL cell is as described in Figure 1 of \cite{Strukov2005-FPGA:}.
These CMOL cells can come in rectilinear fashion as described in 
Fig.~\ref{fig:fpgalogic} but our satifiability formulation 
does not require them to be of any regular shape.

\begin{figure}
\centering
\includegraphics[width=2.5in]{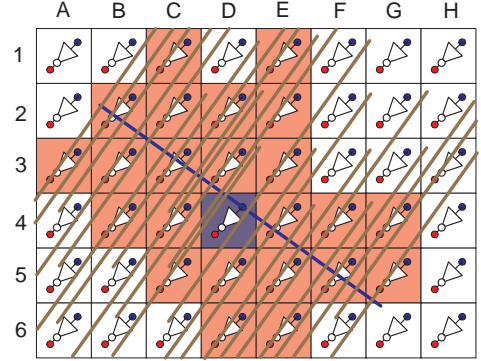}
\caption{An example of connectivity domain without a regular pattern. The input (blue nanowire) of CMOL FPGA cell $D4$ (blue-colored cell) is connected with 23 output nanowires (brown nanowires) from 23 neighbor cells (pink-colored cells).} \label{fig:connect}
\end{figure}

We assume that each CMOL cell $c \in C$ can be connected to a 
set $D(c)$ of CMOL cells, where 
\[D(c) \subset C\]
An example for $D(c)$ is the ``connectivity domain'' as described 
in section~\ref{sec:routing}.
Notice that the ``connectivity domain''
described in \cite{Strukov2005-FPGA:} has a regular pattern around the
neighborhood of each CMOL cell $c$. But the $D(c)$ in our satisfiability
formuation can be arbitrarily any subset of CMOL cells which 
does not require them to have any regular pattern. 
Figure \ref{fig:connect} illustrates such non-regular-patterned 
connectivity domain.

We are given a NOR gate circuit $K$, which can be produced
by Algorithm~\ref{alg:NOR} or other methods.
We can represent the circuit $K$ as a graph $K = (G,E)$ where
$G$ is the set of nodes and $E$ is the set of edges in the graph.
The nodes $G$ corresponds to the gates in the circuit, 
whereas the edges $E$ corresponds to the nets 
(gate-to-gate connections) in the circuit, i.e., $E = G \times G$.
For ease of denotation, we will refer to the gates as $G$ and the
nets as $E$ respectively. Notice that $(g,g') \in E$
if-and-only-if the output of gate $g$ is connected to the input
of gate $g'$ for all gates $g,g' \in G$. We have:
\[\forall g,g' \in G . \{((g,g') \in E) \Leftrightarrow (g \in fanin(g'))\}\]

The CMOL cell assignment problem is to place the circuit $K$ on $C$
such that each gate will occupy one-and-only-one CMOL cell,
and the input-output connectivity of each NOR/NOT gate in the assigned
CMOL cell $c$ falls within the ``connectivity domain'' $D(c)$.
We can describe the CMOL cell assignment problem mathematically 
as an injective function $P$:
\[P : G \rightarrow C\]
where
\begin{eqnarray}
&\forall g,g' \in G . \{(P(g) = P(g')) \Rightarrow (g = g')\}&\\
&\forall g,g' \in G . \{((g,g') \in E) \Rightarrow (P(g) \in D(P(g')))\}&
\end{eqnarray}

We now propose a satisfiability-based (SAT) approach for solving
the CMOL cell assignment problem. We first introduce a set of boolean
variables to encode the various possibilities of placing 
NOR gates to CMOL cells. We then formulate the boolean constraints
to characterize the injective nature of the mapping and the
CMOL requirements.

Let $p_g^c$ be a Boolean variable that represent the assignment of
gate $g$ on CMOL cell $c$, where $g \in G$ and $c \in C$.

Since each gate must be assigned to at most one CMOL cell,
we have:
\begin{equation}
\bigwedge_{g \in G} \left( 
\bigwedge_{c_1,c_2 \in C}^{c_1 \neq c_2} \lnot (p_g^{c_1} \land p_g^{c_2}) 
\right)
\label{eqn:cell_mutex}
\end{equation}

In addition, each gate must be assigned to at least one CMOL cell:
\begin{equation}
\forall g \in G \cdot \exists c \in C \cdot p_g^c\label{eqn:place_exist}
\end{equation}

We cannot assign two or more gates to the same CMOL cell:
\begin{equation}
\bigwedge_{c \in C} \left( 
\bigwedge_{g_1,g_2 \in G}^{g_1 \neq g_2} \lnot (p_{g_1}^c \land p_{g_2}^c) 
\right)
\label{eqn:gate_mutex}
\end{equation}

The connected gates in the circuit must be placed within the
connectivity domain:
\begin{equation}
\bigwedge_{(g_1,g_2) \in E} \left( \bigwedge_{c_2 \in C} \left( 
(\lnot p_{g_2}^{c_2}) \lor \bigvee_{c_1 \in D(c_2)} p_{g_1}^{c_1} 
\right) \right)
\label{eqn:conn_domain}
\end{equation}

We construct a satisfiability formula by conjuncting all the above constraints
(\ref{eqn:cell_mutex}), 
(\ref{eqn:place_exist}), 
(\ref{eqn:gate_mutex}), 
(\ref{eqn:conn_domain}).
We feed all the above constraints to a SAT solver.
The solution that satisfies the conjunction is the cell assignment result.

We can introduce more constraints to address practical issues
of the cell assignment problem. For example, our formulation so far
allows any NOR gate to be assigned to any CMOL cell, as long as
it fits all the above constraints. However, under certain situations,
we want to prevent the assignment of a certain gate $g$ to a subset $C'$
of CMOL cells. To handle such cases,
we simply have to set the boolean variable $p_g^c$ to be FALSE,
where $c \in C'$. We can then propagate constants through the boolean
formulations and simplify the problem. 


\begin{figure}
\centering
\includegraphics[width=2.8in]{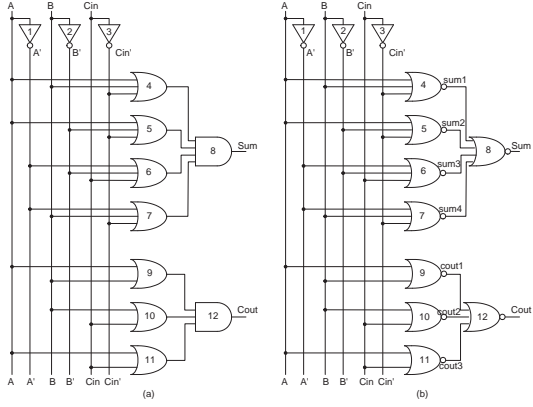}
\caption{Logic circuit for a full adder} \label{fig:adder_nor}
\end{figure}

\begin{figure}
\centering
\includegraphics[width=1.5in]{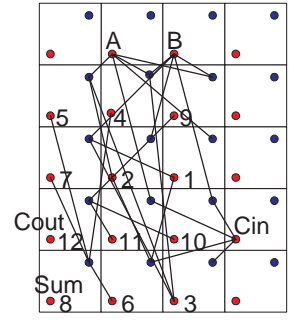}
\caption{CMOL implementation for a full adder} \label{fig:adder_cmol}
\end{figure}

Notice that the cell assignment method presented here is designed to address
technology-specific issues for the CMOL architecture at the lowest
design module level. A top-down hierarchical approach
of both global placement and detailed placement techniques should be
used for large designs.

Contemporary SAT solvers are architected to terminate
early as soon as any solution is found.
If no solution exists, the SAT solver will search all possible cases
and essentially prove to us that the problem is unsolvable.
Notice that if there are more than one solutions to the problem,
the SAT solver tend to finish very fast.
Satisfiability has been used to solve a variety of problems \cite{BLRP:02,segment:TODAES}.
Many commercial CAD tools in industry also use SAT solvers in their CAD flow.

\subsection{Example: Adder}

We first experiment with a simple full adder as an example.
Figure~\ref{fig:adder_nor}(a) shows the logic structure of a full adder
in product of sum format. 
The adder has 3 inputs ($A$, $B$, $Cin$) and 2 outputs ($Sum$, $Cout$).
Using our transformation technique in section~\ref{sec:NOR}, we converted 
the logic circuit to NOR format as shown in figure~\ref{fig:adder_nor}(b).
We then feed the NOR circuit to our CMOL cell assignment tool,
and specified a $4 \times 5$ region with the following restrictions: 
\begin{itemize}
\item inputs $A$ and $B$ must be located at the 
left 3 cells (only 2 are needed)on the top row;
\item input $Cin$ must be located at the right column 
at the second cell from the bottom up;
\item output $Cout$ must be located at the left column
at the second cell from the bottom up (corresponding row with $Cin$);
\item output $Sum$ must be located within the left 3 cells at the bottom row; 
and
\item all other gates must be located within the lower left $3 \times 4$ 
CMOL cell region.
\end{itemize}
All these restrictions are fed to the SAT constraints by setting
the corresponding cell assignment variables to be constant.

The result is shown in figure~\ref{fig:adder_cmol}. 
For each CMOL cell, we use a red dot at the lower-left 
(and a blue dot at the upper-right)
to indicate the output (input) terminals, respectively.
The lines connecting the dots are corresponding to the interconnection
between the NOR gates in figure~\ref{fig:adder_nor}(b). Notice that
these lines are simple logic connection indicators, and should not be
confused with the nanowire crossbar which should be regularly oriented at
an angle relative to the square array of the cells.

\section{Defect Tolerance}\label{sec:defects}

\begin{figure}
\centering
\includegraphics[width=3.2in]{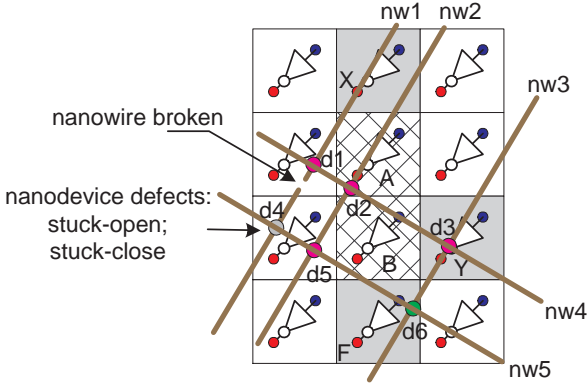}
\caption{Different types of CMOL defects} \label{fig:defect}
\end{figure}

There are many possible causes of defect to the CMOL implementation
as illustrated in figure~\ref{fig:defect}.
In the figure, we use $nw1, nw2, \dots$ to denote the nanowires, 
and $d1, d2, \dots$ to denote the nanodevices, respectively.
The nanodevice $d4$ (colored gray) is defective,
like a pre-programmed ``OFF'' (stuck-open).
The nanodevice $d6$ (colored green) has a different defect,
like a short-circuit.
It is pre-programmed ``ON'' (stuck-closed).
$d1,d2,d3,d5$ are non-defective (colored pink). 
They can be programmed ``ON'' or ``OFF'' by the user.

Given CMOL cells $X$, $Y$, and $F$ in figure~\ref{fig:defect}, 
and assuming there are no defects,
we can implement $F=NOR(X,Y)$ through $nw1,nw3,nw5,d4,d6$.
If nanowire $nw1$ is broken (defect), we have to use CMOL cell $A$
instead of $X$, and program $d5$ ``ON''. 
So $F=NOR(A,Y)$, where $A$ replaces $X$.
If $d4$ is stuck-open, then it doesn't matter $nw1$ is broken or not,
we cannot connect cells $X$ and $F$.
In this case, we can either use $A$ to replace $X$ like above, 
or we can use $B$ to replace $F$, and enable $d1$ and $d3$,
such that $B=NOR(X,Y)$, where $B$ replaces $F$.

In general, we can foresee the following types of defects 
for the CMOL architecture:
\begin{enumerate}
\item A (input/output) nanowire for a CMOL cell is broken into 
two or more segments. Hence the CMOL cell may not be able to
connect to all other $M$ CMOL cells within its input/output connectivity
domain or radius $r$ (where $M=2r(r-1)-1$). In this case,
the CMOL cell is still useful but its connectivity domain
should be modified into a different shape.
\item The nanodevice connecting two perpendicular nanowires 
(let's say the output nanowire of cell A and input nanowire of cell B) 
is stuck-at-open. 
In this case, the connection from A to B through this nanodevice
is broken. But these two CMOL cells can
still be used. We simply have to modify the connectivity domains
such that A is outside the input connectivity domain of B, and
B is outside the output connectivity domain of A.
\item The nano-device connecting two perpendicular nanowires 
(let's say the output nanowire of cell A and input nanowire of cell B) 
is stuck-at-closed. 
In this case, A will always be in the NOR gate input of B. 
To optimize the CMOL cell usage, we have two choices:
\begin{itemize}
\item Do not use cell B, but cell A can still be used if we desire.
\item Assign a NOR gate in cell B, and assign one of B's inputs to cell A.
\end{itemize}
\item Something else is wrong with a CMOL cell randering this cell to be usuable, 
including (but not limited to) the following:
\begin{itemize}
\item the input/output terminal connecting the CMOS layer and the 
input/output nanowire is broken
\item the CMOS inverter is broken
\end{itemize}
\end{enumerate}
Notice that prior work \cite{Strukov2005-FPGA:} is mainly focused on
our defect type 2 above. This paper is the first attempt to address
various other types of defects.

We can formulate the above 4 defects using satisfiability constraints.

For defect types 1 and 2, 
the input/output connectivity domain for the CMOL cells
related to the defect should be modified. Notice that our cell assignment
formulation in section~\ref{sec:satis} does not assume any regularity
for the connectivity domains, so it can be modified to arbitrary shape
depending on the defect.

For defect type 3, the CMOL cell A will always be one of the NOR gate input
for CMOL cell B. We need to make sure that any node $g$ with no fanin
(i.e. primary input) cannot be assigned to cell B. This can be easily
done by setting the corresponding cell assignment variables at cell B to FALSE.
\begin{equation}
p_g^B = 0
\label{eqn:primary-input}
\end{equation}
for all $g \in G$ where $fanin(g) = \emptyset$.

We also need to make sure that any gate assigned to cell B must have
one of its input placed at cell A.
\begin{equation}
(\neg p_{g_1}^B) \lor \bigvee_{g_2 \in fanin(g_1)} p_{g_2}^A
\label{eqn:short-circuit}
\end{equation}
for all $g_1 \in G$ where $fanin(g_1) \ne \emptyset$.

For defect type 4, we cannot assign any gate to the defective CMOL cell.
Hence we must set the cell assignment variable of every gate at that cell to FALSE.
\begin{equation}
p_g^c = 0
\end{equation}
for all $g \in G$, and $c$ is the defective CMOL cell.

Given a manufactured CMOL device with some known defects, and an initial
mapping (most likely done before the manufacturing) that did not take those
defects into account, we need to reconfigure the CMOL device to work
around those defects. 
We can use the following algorithm for the reconfiguration:

\begin{algorithm}
\caption{Reconfiguration}
\label{alg:reconfig}
\begin{algorithmic}[1]
\STATE find all assigned cells that conflicts with the defects
\REPEAT
\STATE compute the center of mass of all conflicts
\STATE cut a small region $R$ around the center of mass
\REPEAT
\STATE enlarge the region $R$
\STATE redo cell assignment in $R$ to avoid conflict
\UNTIL{cell assignment is successful}
\UNTIL{no more conflicts}
\end{algorithmic}
\end{algorithm}

Our algorithm takes advantage of the idea that many defects in the real world
tend to be clustered together. So we use a center of mass computation
to focus our reconfiguration region on the affected cells.

\section{Experiments}
\label{sec:exp}



Besides the adder example, we conducted experiments on ISCAS benchmarks.
For each design, we first carve out the combinational logic,
i.e., convert all inputs/outputs of latches (or sequential elements)
to primary outputs/inputs respectively. We then run the SIS sweep operation
to simplify the circuit (and remove redundant gates).
We convert the circuit to NOR gates using the simple algorithm
in Section~\ref{sec:NOR}. To make sure that our logic transformations
are correct, we formally verified the logic correctness
of the transformation results using equivalence checkers. 
Finally, we assign the NOR gates to CMOL cells
using our method in Section~\ref{sec:satis}. 
The CMOL cell region for each design was choosen to be a square shaped 
(or nearly square) territory where the number of cells is slightly
more than the number of NOR gates in the circuit. We also put
special restriction so that the primary inputs/outputs are located
around the perimeter of the square region.
We use the connectivity domain with radius $r = 9$
as specified in \cite{Strukov2005-FPGA:}.
We ran our satisfiability constraints through the 
MiniSAT~\cite{minisat:JSAT2006} solver to generate the cell assignment.
All experiments are performed on Linux with a 1.5GHz Intel Pentium 4 CPU
with 3GB memory.
Table~\ref{expiscas} summarizes our experiments.
The inputs/outputs column shows the primary inputs/outputs after carving 
out the combinational part of the benchmark design.
The cells column shows the number of NOR/NOT gates that need to be assigned
to CMOL cells. The X and Y dimensions show the size of the cell assignment
territory. The vars and clauses indicate the size of the CNF
generated for the satisfiability formulation.
The time column indicates the CPU seconds that the MiniSAT took
to solve the problem.

\begin{table*}
\centering
\caption{\label{expiscas}Cell Assignment Experiments}
\begin{tabular}{|l|r|r|r|r|r|r|r|r|}
\hline
Circuit & inputs & outputs & cells & X & Y & vars & clauses & time (sec)\\
\hline
adder & 3 &  2 &  12 &  4 &  5 &   135 &     1577 & 0.01\\
s208 & 18 &  9 & 136 & 13 & 12 & 18246 & 2444373  & 509.84\\
s27  &  7 &  4 &  19 &  5 &  5 &   376 &    7164  &   0.07\\
s298 & 17 & 20 & 122 & 13 & 12 & 14962 & 1829223  & 370.30\\
s344 & 24 & 26 & 179 & 14 & 14 & 27884 & 4595793  &   6.18\\
s349 & 24 & 26 & 184 & 14 & 14 & 28864 & 4834296  &   7.60\\
s382 & 24 & 27 & 175 & 14 & 14 & 26956 & 4377249  &  12.88\\
s386 & 13 & 13 & 164 & 14 & 13 & 26416 & 4278318  &  10.30\\
s400 & 24 & 27 & 188 & 15 & 14 & 31524 & 5540402  &   7.52\\
s444 & 24 & 27 & 187 & 15 & 14 & 31314 & 5489242  &   7.59\\
s510 & 25 & 13 & 266 & 18 & 18 & 88768 & 26315936 & 213.27\\
\hline
\end{tabular}
\end{table*}

For defect tolerance, we inject the various defects mentioned in
section~\ref{sec:defects} to the cell assignments in Table~\ref{expiscas}.
For each design, we first pick a randomly generated location $(x_0,y_0)$,
and then compute a probability density function $pdf(x,y)$ for each
location using the Gaussian distribution.
\[pdf(x,y) = \frac{1}{\sigma\sqrt{2\pi}}e^{-\frac{(x-x_0)^2 + (y-y_0)^2}{2\sigma^2}}\]
This probability density function is used to control the injection of defects.
For each defect type, we generate a random floating point number $rand()$
between 0 and 1 at each location where the defect can happen. 
The defect is injected if $rand() \leq pdf(x,y)$.
The cumulative SAT solver runtime that Algorithm~\ref{alg:reconfig}
took for each design is shown in Table~\ref{tab:reconfig}.

\begin{table}
\centering
\caption{\label{tab:reconfig}Reconfiguration Experiments}
\begin{tabular}{|l|r|r|}
\hline
Circuit & $\sigma = \frac{r}{3}$ & $\sigma = \frac{2r}{3}$\\
\hline
s208 & 0.26 & 0.12\\
s298 & 0.11 & 0.42\\
s344 & 0.21 & 0.31\\
s349 & 0.26 & 0.45\\
s382 & 0.18 & 0.25\\
s386 & 0.14 & 4.00\\
s400 & 0.02 & 0.61\\
s444 & 0.01 & 0.48\\
s510 & 0.12 & 0.06\\
\hline
\end{tabular}
\end{table}


\section{Conclusion}
\label{sec:conclude}

In this paper, we present a CAD framework for the CMOL architecture.
We transform any netlist of AND/OR/NOT gates to a NOR gate circuit, 
and then map the NOR gates to CMOL cells.
Our CMOL cell assignment is based on satisfiability
and it can generate the assignment if and only if the solution exists.
We also present a model for various types of defects under the CMOL
architecture, most of which has never been studied before. 
We present a reconfiguration technique that can
deal with all these defects through our CAD framework.
This is the first work on automated CMOL cell assignment,
and the first to model and tolerate several different CMOL defects.
Our experiments indicate that our reconfiguration technique is fast
and scalable.


\begin{thebibliography}{10}\setlength{\itemsep}{-1ex}\small

\bibitem{Bachtold2001-Logic}
A.~Bachtold, P.~Hadley, T.~Naknishi, and C.~Dekker.
\newblock Logic circuits with carbon nanotube transistors.
\newblock {\em Science}, 294:1317--1320, 2001.

\bibitem{Chen2003-Nanoscale}
Y.~Chen, G.-Y. Jung, D.~A.~A. Ohlberg, X.~Li, D.~R. Stewart, J.~O. Jeppesen,
  K.~A. Nielsen, J.~F. Stoddart, and R.~S. Williams.
\newblock Nanoscale molecular-switch crossbar circuits.
\newblock {\em Nanotechnology}, 14:462--468, 2003.

\bibitem{DeHon2003-Stochastic}
A.~DeHon and e.~al.
\newblock Stochastic assembly of sublithographic nanoscale interfaces.
\newblock {\em IEEE Trans. on Nanotechnology}, 2:165--174, 2003.

\bibitem{minisat:JSAT2006}
N.~Een and N.~Sorensson.
\newblock {Translating Pseudo-Boolean Constraints into SAT}.
\newblock {\em Journal on Satisfiability, Boolean Modeling and Computation},
  2:1--26, 2006.

\bibitem{Friedman2005-High-speed}
R.~S. Friedman, M.~C. McAlpine, D.~S. Ricketts, D.~Ham, and C.~M. Lieber.
\newblock High-speed integrated nanowire circuits.
\newblock {\em Nature}, 434(28):1085, 2005.

\bibitem{Gao2007-Cortical}
C.~Gao and D.~Hammerstrom.
\newblock Cortical models onto {CMOL} and {CMOS} - architectures and
  performance/price.
\newblock {\em submitted to IEEE Tran. on Circuits and Systems - I}, 2007.

\bibitem{segment:TODAES}
W.~N.~N. Hung, X.~Song, E.~M. Aboulhamid, A.~Kennings, and A.~Coppola.
\newblock {Segmented Channel Routability via Satisfiability}.
\newblock {\em ACM Transactions on Design Automation of Electronic Systems},
  9(4):517--528, October 2004.

\bibitem{Kuekes2005-crossbar}
P.~J. Kuekes, D.~R. Stewart, and R.~S. Williams.
\newblock The crossbar latch: logic value storage, restoration, and inversion
  in crossbar circuits.
\newblock {\em Journal of Applied Physics}, 97:034301--1--5, 2005.

\bibitem{Likharev2005-CMOL:}
K.~K. Likharev and D.~V. Strukov.
\newblock {CMOL}: devices, circuits, and architectures.
\newblock In G.~Cuniberti and et~al., editors, {\em Introduction to Molecular
  Electronics}, pages 447--477. Springer, Berlin, 2005.

\bibitem{Resnick2003-Imprint}
D.~J. Resnick, W.~J. Dauksher, D.~Mancini, K.~J. Nordquist, T.~C. Bailey,
  S.~Johnson, N.~Stacey, J.~G. Ekerdt, C.~G. Willson, and S.~V. Sreenivasan.
\newblock Imprint lithography for integrated circuit fabrication.
\newblock {\em Journal of Vacuum Science {\&} Technology B: Microelectronics
  and Nanometer Structures}, 21:2624, 2003.

\bibitem{Snider2007-Nano/CMOS}
G.~Snider and R.~Williams.
\newblock Nano/cmos architectures using a field-programmable nanowire
  interconnect.
\newblock {\em Nanotechnology}, 18:1--11, 2007.

\bibitem{Snider2004-CMOS-like}
G.~S. Snider, P.~J. Kuekes, and R.~S. Williams.
\newblock {CMOS}-like logic in defective, nanoscale crossbars.
\newblock {\em Nanotechnology}, 15:881--891, 2004.

\bibitem{BLRP:02}
X.~Song, W.~N.~N. Hung, A.~Mishchenko, M.~Chrzanowska-Jeske, A.~Coppola, and
  A.~Kennings.
\newblock Board-level multiterminal net assignment for the partial cross-bar
  architecture.
\newblock {\em IEEE Transactions on VLSI Systems}, 11(3):511--514, June 2003.

\bibitem{Strukov2006-reconfigurable}
D.~Strukov and K.~Likharev.
\newblock A reconfigurable architecture for hybrid {CMOS}/nanodevice circuits.
\newblock In {\em FPGA'06}, pages 131--140, Monterey, California, USA, 2006.

\bibitem{Strukov2005-FPGA:}
D.~B. Strukov and K.~K. Likharev.
\newblock {CMOL FPGA}: a reconfigurable architecture for hybrid digital
  circuits with two-terminal nanodevices.
\newblock {\em Nanotechnology}, 16(6):888--900, 2005.

\bibitem{Strukov2005-Prospects}
D.~B. Strukov and K.~K. Likharev.
\newblock Prospects for terabit-scale nanoelectronic memories.
\newblock {\em Nanotechnology}, 16:137--148, 2005.

\bibitem{Turel2005-Architectures}
O.~T\"urel, J.~H. Lee, X.~Ma, and K.~K. Likharev.
\newblock Architectures for nanoelectronic implementation of artificial neural
  networks: new results.
\newblock {\em Neurocomputing}, 64:271--283, 2005.

\bibitem{Xiang2006-Ge/Si}
J.~Xiang, W.~Lu, Y.~Hu, Y.~Wu, H.~Yan, and C.~M. Lieber.
\newblock Ge/{S}i nanowire heterostructures as high-performance field-effect
  transistors.
\newblock {\em Nature}, 441(25):489--493, 2006.

\bibitem{Zankovych2001-Nanoimprint}
S.~Zankovych, T.~Hoffmann, J.~Seekamp, J.-U. Bruch, and C.~M.~S. Torres.
\newblock Nanoimprint lithography: challenges and prospects.
\newblock {\em Nanotechnology}, 12:91--95, 2001.

\bibitem{Ziegler2003-CMOS/nano}
M.~M. Ziegler and M.~R. Stan.
\newblock {CMOS}/nano co-design for crossbar-based molecualr electronic
  systems.
\newblock {\em IEEE Transactions on Nanotechnology}, 2(4):217--230, 2003.

\end{thebibliography}
\end{document}